\documentclass[preprint,double-spaced,aps,apssymb]{revtex4}
\usepackage{amsmath,amssymb}
\usepackage{graphicx}
\usepackage{dcolumn}
\usepackage{bm}
\usepackage{subfigure}
\newcommand{\ber}{\begin{eqnarray}}
\newcommand{\eer}{\end{eqnarray}}
\newcommand{\bea}{\begin{equation}}
\newcommand{\eea}{\end{equation}}

\begin{document}

\title{Does the fluctuation-dissipation relation guarantee equilibrium?}

\author{A. Bhattacharyay}
\email{a.bhattacharyay@iiserpune.ac.in}
 \affiliation{Indian Institute of Science Education and Research, Pune, India}

\date{\today}

\begin{abstract}
We study a symmetry broken harmonic oscillator in contact with a heat bath characterized by a fixed temperature. The overdampped system is solved exactly to show symmetry broken directed transport raising the question whether fluctuation-dissipation relation does guarantee equilibrium.   
\end{abstract}
\maketitle

Consider two bodies of same mass but of different damping (due to difference in shape size etc) are connected by a spring and the system is in contact with a heat bath of temperature T. The equations of motion are the following
\ber
\frac{\partial^2x_1}{\partial t^2}=(1-\beta)\frac{\partial x_1}{\partial t}-\alpha(x_1-x_2)+\sqrt{2(1-\beta)k_BT}\eta_1 -V\frac{\partial \Theta (x_2-x_1)}{\partial x_1}\\
\frac{\partial^2x_2}{\partial t^2}=\frac{\partial x_2}{\partial t}-\alpha(x_2-x_1)+\sqrt{2k_BT}\eta_2 -V\frac{\partial \Theta (x_2-x_1)}{\partial x_2}
\eer 
In the above system the coordinates $x_1$ and $x_2$ are the center of mass of the bodies 1 and 2 where $x_1>x_2$ and the hard core volume exclusion would be applied (later) by taking the limit V $\rightarrow $ infinity. The spring constant is $\alpha$, $<\eta_i>=0$ and $<\eta_i(t)\eta_j(s)>=\delta_{ij} \delta(t-s)$ ensures the system to be at a temperature T according to the standard fluctuation dissipation relationship where $k_B$ is the Boltzmann constant.  

Taking the overdamped limit and moving to the internal coordinate $Z=x_1-x_2$ and centre of mass $X=\frac{x_1-x_2}{2}$ we can re write the above system as
\ber
\dot{Z}=-\frac{\alpha(2-\beta)}{1-\beta} Z+ \xi_Z -\frac{V(2-\beta)}{1-\beta}\frac{\partial \Theta (-Z)}{\partial Z}\\
\dot{X}=-\frac{\alpha\beta}{2(1-\beta)} Z+\xi_X + \frac{V\beta}{2(1-\beta)}\delta(Z).
\eer
Here $<\xi_Z(t)\xi_Z(s)>=2Tk_B\frac{2-\beta}{1-\beta}\delta(t-s)$ and $<\xi_X(t)\xi_X(s)>=Tk_B\frac{2-\beta}{2(1-\beta)}\delta(t-s)$

The distribution of Z follows from Eq.3 as 
\bea
P(Z)=\exp{(-\frac{\alpha Z^2}{2k_BT}-\frac{V\Theta (-Z)}{k_BT})}
\eea
The hardcore repulsion term at the limit V tending to infinity will suppress the distribution on the negative Z space and that will result in a nonzero $<Z>$. The average velocity of the cm being $<\dot{X}>=-\frac{\alpha\beta}{2(1-\beta)}<Z> + \frac{V\beta}{2(1-\beta)}\int_\epsilon^\infty{\delta(Z)P(z)dz}$ will now be nonzero where $\epsilon$ is an infinitesimal positive constant. The second term in the expression for $<\dot{X}>$ is zero for a nonzero $\epsilon$ where $\epsilon=0$ is not accessible because of excluded volume i.e. the Heaviside step function $\Theta(-Z)$ or equivalently by the vanishing of the $P(Z)$ at $Z\leq 0$.
\par
Let us note that no contribution from the second term in the expression of $<\dot{X}>$ may be understood from a mechanistic view point. All the interactions between two bodies under consideration such as hard core collision or spring interaction are themselves momentum preserving. The resultant motion of the CM is due to different damping experienced by two bodies on the flight due to the interaction with the environment. Thus, the requirement according to the Newton's first law is fulfilled where there is an external force (damping) that causes nonzero velocity of the CM of the system. Now, the hard core collision (infinite V) is ideally a momentary event when the velocity of the colliding particles are zero. Just before and after the collision the particles have velocities. So, the hard core collision should remain a momentum preserving event and as a consequence there should not be any contribution to the velocity of the CM from the hard core collision which justifies the vanishing of the contribution to $<\dot{X}>$ from the hard core collision term. Another point to note that, in the above calculations, the qualitative results are practically independent of the strength of the thermal noise ($\eta_i$) so long as $\eta_1$ and $\eta_2$ are uncorrelated (which practically led to the presented calculations from a previously otherwise considered model \cite{ari}). To bring such a system to equilibrium would there be a correlation established in $\eta_is$? If so, it goes somewhat against the existing concept of representation of a heat bath in equilibrium.
\par
This is something extraordinary! Extraordinary because the system is not subjected to any external force simply the symmetry has been broken by the difference in the damping constants and the strength of the Gaussian noise being proportional to the damping constants keeps the temperature seen by the two bodies making up the system constant. In the standard form of a Langevin equation there always is the scope to keep an external force felt by the particle and a system can always move in a constant temperature heat bath under the action of this external force showing some fluctuations on its deterministic path of motion. But here, there is an internal field felt by the system and still it moves being in contact with a heat bath characterized by a constant temperature as per the demand of the fluctuation-dissipation relation. But, in equilibrium a system cannot move directionally which is in conflict with the concept of the stationary probability distribution. So, the question naturally arises as to what extent the derivation of the fluctuation-dissipation relation for a single particle in equilibrium with a heat bath applies for a system of particles having an internal time scale ($1/\alpha$). In the derivation of the FDT its generally taken into account that the heat bath equilibrates very quickly \cite{reif} and that too without any reference to a particular time scale with respect to which it should be quickly. But when, there is a particular characteristic time scale involved should it be defined differently if we assert that there is nothing wrong with the overdamped limit?  
\section {Zwanzig model}
Let us derive the model Eq.1 and Eq.2 following the method of Zwanzig's to show that under certain conditions where the couplings of the two bodies with the environmental degrees of freedom are uncorrelated the cross terms (in momentum) in Eq.1 and 2 can be avoided and Eq.1 and 2 retain their shape and do not become a standard Ornstein-Uhlenbeck one.
Consider the Hamiltonian of the system as 
\bea
H=\frac{P_1^2}{2}+\frac{P_2^2}{2}+V(X_1,X_2)+\sum_1{\frac{\omega_i^2}{2}\left( q_i-\frac{\gamma_{1i}X_1}{\omega_i^2}\frac{\gamma_{2i}X_2}{\omega_i^2}-\right )^2} +\sum_i{\frac{p_i^2}{2}}.
\eea
In the above expression, $q_is$ are the bath degrees of freedom. The dynamics that results from the above mentioned equation are
\ber
\dot{P_1}=-\frac{\partial V(X_1,X_2)}{\partial X_1} + \sum_i{\gamma_{1i}\left( q_i-\frac{\gamma_{1i}X_1}{\omega_i^2}\frac{\gamma_{2i}X_2}{\omega_i^2}-\right )}\\
\dot{P_2}=-\frac{\partial V(X_1,X_2)}{\partial X_2} + \sum_i{\gamma_{2i}\left( q_i-\frac{\gamma_{1i}X_1}{\omega_i^2}\frac{\gamma_{2i}X_2}{\omega_i^2}-\right )}
\eer
and
\bea
\dot{p_i} + \omega_i^2q_i = \gamma_{1i}X_1 + \gamma_{2i}X_2
\eea

Following the standard procedure Eq.7 and 8 can be rewritten as
\ber
\dot{P_1} = -\frac{\partial V}{\partial X_1} - \sum_i{\int_0^t\frac{\gamma_{1i}^2}{\omega_i^2}\cos{\omega_i(t-s)}P_1(s)ds} -\sum_i{\int_0^t\frac{\gamma_{1i}\gamma_{2i}}{\omega_i^2}\cos{\omega_i(t-s)}P_2(s)ds} +f_1(t)\\
\dot{P_2} = -\frac{\partial V}{\partial X_2} - \sum_i{\int_0^t\frac{\gamma_{2i}^2}{\omega_i^2}\cos{\omega_i(t-s)}P_2(s)ds} -\sum_i{\int_0^t\frac{\gamma_{1i}\gamma_{2i}}{\omega_i^2}\cos{\omega_i(t-s)}P_1(s)ds} +f_2(t)
\eer

In the above equations $f_1(t)$ and $f_2(t)$ are initial condition dependent. Since there is a sum over the bath degrees of freedom in the cross terms containing $\gamma_{1i}\gamma_{2i}$, under the condition that the $\gamma_{1i}$ and $\gamma_{2i}$ are uncorrelated this cross term should vanish making the equations 10 and 11 take the shape of Eq.1 and 2. So, the form of Eq.1 and 2 makes sense where we have not taken into account the cross terms as is done in a standard Ornstein-Uhlenbeck process. Now, the results shown above are a consequence of the structure of Eq.1 and 2 and that holds at least under the situation when the coupling of the two bodies (connected by spring) to the environmental degrees of freedom is uncorrelated. 
\par
Zwanzig model although is a widely accepted microscopic picture leading to the Langevin dynamics. However, its physically far from reality because the long range strong coupling between all the degrees of freedoms. Nevertheless, it at least, when put in the context of the introductory section's results, indicates that the uncorrelated coupling of the system with the degrees of freedom of the bath is something leading to the breakdown of second law here and breakdown of second law is something totally unacceptable. So, the indication made in the introductory section that to restore equilibrium i.e. restore second law a correlated forcing of the system would be necessary is at least getting some support following this Zwanzig-model calculations.
\par
If we think physically, how would the equilibrium be established? Consideration of inherent uniformity (or symmetry) of an equilibrium state when put together with the symmetry broken dimer is leading to all these problems. So, if the dimer is symmetry broken (shape and size of the two bodies are different) then their local atmosphere will also get symmetry broken by having variations in density of particles etc. This is a situation where the change in the local environment of the two bodies making the dimer would restore the broken symmetry making the damping of two very different objects the same. Now, this whole locally varied environment of the dimer has to come to terms with the global uniform (as equilibrium normally demands) state and that remains an issue. The possibility of a global symmetry breaking cannot be ruled out even in equilibrium and probing its dynamics since the insertion of the dimer into an uniform equilibrium system might be interesting.    
\section{Non-singular collision potential}
It may appear that the result shown above is a consequence of a hard core potential only and to counter that in this section we would consider a nonsingular potential for the collision of the two bodies constituting the dimer \cite{udo}. Instead of considering the $\Theta(-Z)$ a Heaviside step function, let us take the potential as $\Theta(Z)= -a(z-z_0) for Z\leq Z_0$ and $\Theta(Z)=0 for Z>Z_0$. The discontinuity at the point $Z_0$ is essential as the constituents of the dimer are just coming in contact with each other at this gap. Moreover, we would further consider that the gradient $a$ of this repulsive potential is steep enough to ensure non-passage of the classical particles past each other to make $Z$ negative. So, $a$ and are adjustable parameters at our hand. With such a potential the dynamics on internal coordinate and the CM now takes the form
\ber
\dot{Z}=-\frac{\partial}{\partial Z}\left [ \frac{\alpha(2-\beta)}{2(1-\beta)}Z^2-\frac{2-\beta}{1-\beta}a(Z-Z_0)_{for Z \leq Z_0}  \right ]\\
\dot{X}=-\frac{\alpha\beta}{2(1-\beta)}Z+\frac{\beta}{2(1-\beta)}a_{for Z\leq Z_0}
\eer
\par
The probability distribution for the Z would be
\bea
P(Z) = \exp{\left( -\frac{\alpha Z^2}{2k_B T}+\frac{a(Z-Z_0)}{k_BT}_{for Z\leq Z_0} \right )}.
\eea
Now, we can write the average velocity of the CM as
\bea
<\dot{X}>=-<\frac{\alpha\beta Z}{2-\beta}\left [e^{\frac{a(Z-Z_0)}{k_BT}_{for Z\leq Z_0}}\right ]> +<\frac{\beta a}{2(1-\beta)} \left [e^{\frac{a(Z-Z_0)}{k_BT}_{for Z\leq Z_0}}\right ]>
\eea
The averages on the right hand side are now for the Gaussian probability distribution $\exp{\left( -\frac{\alpha Z^2}{2k_B T}\right )} $ only. Definitely these two terms on the right hand side of the above equation would not cancel in general being the averages of two very different quantities with a lots of free parameters involved. Note that, by the consideration of $a$ being large enough we are doing away with any probability of $Z<0$ and the integration limit is practically from $0$ to $\infty$ while averaging. Thus, any objection considering the delta function singularity (or equivalently infinitely discontinuous potential) only resulting in the CM velocity does not stand. So long as the excluded volume is there the result is there.
\par
Let us try to understand why the resultant velocity of the CM is practically independent of the type of the repulsive potential. The repulsive excluded volume potential and the spring potential can actually be two very different things giving the scope of having many tunable free parameters to do away with any cancelation of the positive and negative parts of the CM velocity resulting in zero resultant velocity. Physically, it means that you can always control the range of the flight of the two bodies making the dimer before they collide and how they collide and the external forcing resulting in the CM motion practically works during the flight. Thus, issue whether the fluctuation-dissipation relation guaranties equilibrium, now, more certainly does indicate that it does not always do that.
\section{How to extract energy}
In order to design an engine to extract energy from a quasi one dimensional bath one has to simply introduce sush an asymmetric dimer into the system. As soon as the dimer is introduced the system turns nonequilibrium and starts its journey to the next equilibrium state where the density of the environment would be different on the two sides of the dimer. Actually, one can imagine that the dimer moves to achieve this density varied state by changing the effective volume on either sides of it. So, long as the required density veriation is achieved the dimer keeps moving to adjust the volume. Note that, here we are considering the quasi 1D heat bath is in contact with a bigger heat bath (atmosphere say) to keep its temperature constant. Now, once the dimer has stopped following reverse density gradient being established which is strong enough to stop it, one simply has to remove the dimer and allow the system to come back to the initial state. If the energy cost in the introduction and removal of the dimer is less than the energy extracted from its motion there is always a gain in energy. Most important to note that here one is extracting the energy directly from the bath unlike in Brownian ratchet models where work is done by the external forcing. So, the proposed mechanism is a completely novel one and is in contrast with the standard paradigm of Brownian Ratchetting.
\section{Ornstein-Uhlenbeck structure}
The model we are considering here in 2D differs from the structure of a standard Ornstein-Uhlenbeck (OU) form in the sense that it does not include the cross terms of velocities in the eq.1 and eq.2. A possible justification for that has been given in the section Zwanzig-model. Here we will note that there are certain problems with the OU form of this model as well because of the singular nature of the matrix of the elastic force. The standard OU form for our model is
\bea
{\begin{pmatrix} 1-\beta & D_1 \\ D_2 & 1 \end{pmatrix}}{\begin{pmatrix} \dot{x_1} \\ \dot{x_2}\end{pmatrix}}={\begin{pmatrix}-\alpha & \alpha \\ \alpha & -\alpha  \end{pmatrix}}{\begin{pmatrix} x_1 \\ x_2 \end{pmatrix}} + {\begin{pmatrix} \eta_1 \\ \eta_2 \end{pmatrix}}
\eea 
where we have considered hardcore collision potential of practically no width (i.e delta forcing at collision). This can be taken into the form of a standard OU process as
\bea
\dot{X_i}+\sum_i^2{\Gamma_{ij}X} = \eta_{ij}
\eea 
where 
\bea
\Gamma = -{\begin{pmatrix} 1-\beta & D_1 \\ D_2 & 1 \end{pmatrix}}^{-1}{\begin{pmatrix}-\alpha & \alpha \\ \alpha & -\alpha  \end{pmatrix}}
\eea
\par
One can readily find out the eigenvalues of the matrix $\Gamma$ which are 
\ber
\lambda_1 &=& 0 \\
\lambda_2 &=& \frac{\alpha(2+D_1+D_2-\beta)}{1-\beta -D_1D_2}.
\eer
The zero eigenvalue is definitely a problem here because for the system to reach a stationary state the eigenvalues have to be greater than zero. The surprising thing is that in this whole calculation we are not considering a strong external drive which basically takes the system out of the linear response regime where the fluctuation-dissipation relation does not in any way hold. Any strong interaction the system encounters apart from that with the bath degrees of freedom is through the collision of the two bodies to ensure volume exclusion and that is something pretty much naturally occuring everywhere. If this collision is taking the system out of the linear response regime then it shoudl always happen and the credibility of the fluctuation-dissipation relation in its standard form is questionable.

\end{document}